        \newcommand{\DOT}{\hspace{-0.08in}{\bf .}\hspace{0.1in}}
        \newcommand{\BOX}{\hbox {$\sqcap$ \kern -1em $\sqcup$}}
        \newcommand{\qed}{\hskip 3em \hbox{\BOX} \vskip 2ex}
        \newtheorem{question}{Question}
        
        \newtheorem{theorem}{Theorem}
        \newtheorem{lemma}{Lemma}
        
        \def\R{{\bf R}}
        \def\C{{\bf C}}
        \def\Z{{\bf Z}}
        \newcommand{\be}{\begin{equation}}
        \newcommand{\ee}{\end{equation}}
        
        \newcommand{\ba}{\begin{eqnarray}}
        \newcommand{\eps}{\epsilon}
        
        \newcommand{\om}{\omega}

        \newcommand{\maps}{\colon}
        \newcommand{\ea}{\end{eqnarray}}

        \renewcommand{\L}{{\cal L}}
        
        \newcommand{\A}{{\cal A}}
        \newcommand{\G}{{\cal G}}
        \newcommand{\E}{{\cal E}}
        
        \renewcommand{\Re}{{\rm Re}}
        \renewcommand{\Im}{{\rm Im}}

        \newcommand{\tr}{{\rm tr}}
        
        \newcommand{\Diff}{{\rm Diff}}
        \newcommand{\Hom}{{\rm Hom}}
        \newcommand{\Aut}{{\rm Aut}}
        \newcommand{\supp}{{\rm supp}}
        \newcommand{\torus}{S^1 \times D^{n-1}}
        \newcommand{\Tubes} {{\bf T}}
        \renewcommand{\H} {{\bf H}}

        \documentstyle[12pt]{article}
\begin{document}

        \begin{center}
{\bf Link Invariants, Holonomy Algebras and Functional Integration\\}
        \vspace{0.5cm}
        {\em John C. Baez\\}
        \vspace{0.3cm}
        {\small Department of Mathematics \\
        University of California\\
        Riverside CA 92521\\}
        \vspace{0.3cm}
        {\small December 18, 1992}
        \vspace{0.5cm}
        \end{center}

\begin{abstract}
Given a principal $G$-bundle over a smooth manifold $M$, with $G$ a
compact Lie group, and given a finite-dimensional
unitary representation $\rho$ of $G$, one may define an algebra of
functions on $\A/\G$, the ``holonomy Banach
algebra'' $\H_b$, by completing an algebra generated by regularized
Wilson loops.  Elements of the dual $\H_b^\ast$ may be regarded as
a substitute for measures on $\A/\G$.
 There is a natural linear map from $\Diff_0(M)$-invariant
elements of $\H_b^\ast$ to the space of complex-valued ambient isotopy
invariants of framed oriented links in $M$.  Moreover, this map is
one-to-one.  Similar results hold for a C*-algebraic analog, the
``holonomy C*-algebra.''
\end{abstract}

\section{Introduction}

Our goal in this paper is to provide a rigorous expression of some ideas
from physics concerning the relation of link invariants to
diffeomorphism-invariant states of gauge field theories.
In quantum field theory it is common to work formally with ``measures''
on the space of field configurations, ignoring the difficulty of
describing these measures in mathematically rigorous terms.
The simplest example arises in the theory of free boson fields.
Here one treats ``measures'' of the form
\[       e^{Q(x,x)} dx   \]
where $Q$ is a complex-valued symmetric bilinear form on the real vector
space $V$.  When $V$ is finite-dimensional and $dx$ denotes Lebesgue
measure, $e^{Q(x,x)} dx$ is a Borel measure
on $V$.  When $V$ is infinite-dimensional, there is no Lebesgue measure
on $V$, but one may still seek some reasonable way to make sense of
the expression
\[         \int_V  f(x)  e^{Q(x,x)} dx , \]
at least for some class of complex-valued functions $f$ on $V$.

One approach is as follows.
Let $\E$, the {\it exponential algebra,} be the algebra of functions on
$V$ generated by those of the form
\[ f(x) =  e^{Q(v,x)}  \]
for  $v \in V$.
Note that if $V$ is finite-dimensional and
${\rm Re}\,Q(x,x) < 0$ for all nonzero $x \in V$ we have
$$ {\int_{V}  e^{Q(v,x)} e^{Q(x,x)}\, dx \over
\int_{V} e^{Q(x,x)}\, dx   } = e^{Q(v,v)/4}\; $$
Generalizing from this fact, for an arbitrary vector space
$V$ and quadratic form $Q$ one may simply define a
complex-linear functional $E \colon \E \to \C$, often called the
``expectation value,'' such that
$$E\left(e^{Q(v,\cdot)} \right) =  e^{Q(v,v)/4 }\;.$$

The case where $Q$ is real-valued and negative definite
has been widely used in constructive quantum field theory \cite{BSZ}.
In this case the relation of the expectation value to
conventional integration theory is more than a merely formal one.
Since
$$ E(1) = 1, $$
$$ f \ge 0\; \Rightarrow \; E(f) \ge 0,$$
and
$$f \ge 0 \;  \Rightarrow\;  |E(fg)| \le  \|g\|_{\infty} E(f) ,$$
a theorem of Segal says that $\cal E$ may be embedded as a dense subspace
of $L^1(X)$ for a probability measure space $X$
(unique up to a suitable equivalence
relation) on which $E$ is represented as integration \cite{SK}.
Moreover, one can complete $\E$ in the $L^\infty$ norm to obtain a
C*-algebra $\E_c$, and the expectation value extends uniquely to a
continuous linear functional $E \maps \E_c \to \C$.

On the other hand,
Feynman integrals, in which $Q$ is imaginary, arise frequently in
practical calculations in quantum field theory.
The question of extending the expectation value $E \maps \E \to \C$ to a
larger algebra of functions on $V$
been studied extensively in this case \cite{AHK,CS,Johnson}.
In this case, $E \maps \E \to \C$ does not extend
to a continuous linear functional on the completion of $\E$ in the
$L^\infty$ norm, as
is easily checked in the case when $V = \R$ and $Q(v,v) = iv^2$,
where the $L^\infty$ completion of $\E$ is the space of almost periodic
functions.   Since $L^\infty$ norm is the unique
C*-norm on $\E$,
it follows that $\E$ admits no C*-algebra completion to which
$E$ extends to a continuous linear functional.
One can, however, find a Banach algebra completion
of $\E$ to which $E$ extends continuously.  For example, one can
complete $\E$ in the norm given by
\[           \|\sum_i c_i e^{Q(v_i,x)} \| = \sum_i |c_i|   \]
if the points $v_i$ are distinct.

In this paper we consider an extension of this idea from
the linear context of the free boson field to nonabelian
gauge field theories.  An interesting example of a
diffeomorphism-invariant gauge field theory is Chern-Simons theory.
Given a compact Lie group $G$ and a trivial principal $G$-bundle $P \to
M$, the Chern-Simons functional of a connection $A$ on $P$ is given by
\[   S(A) = {k\over 4\pi} \int_{M} \tr (A \wedge dA + {2\over 3}A \wedge A
\wedge A) ,\]
where the level $k \ge 0$ is an integer.   The quantity $S(A)$ changes
by an integer multiple of $2\pi$ under gauge transformations of $A$, so
one may formally define a ``measure''
\[          e^{iS(A)} {\cal D}A \]
on the space $\A/\G$ of smooth connections modulo gauge transformations,
where ${\cal D}A$ is the (purely heuristic) ``Lebesgue measure'' on
$\A/\G$. Since $S(A)$ is invariant under the action under the group
$\Diff_0(M)$ of diffeomorphisms of $M$ that are connected to the
identity, at a formal level we expect the Chern-Simons ``measure'' to be
$\Diff_0(M)$-invariant.

The notion of the holonomy around a loop provides an interesting
relation between link invariants
and such diffeomorphism-invariant ``measures.''
Given an oriented piecewise smooth loop $\gamma \maps S^1 \to M$, let
$T(\Gamma,A)$ denote the trace of the holonomy of the connection $A$
around $\gamma$, relative to some finite-dimensional unitary
representation of
$G$.   The function $T(\gamma) = T(\gamma, \cdot)$ is called a
{\it Wilson loop} in physics.  (For an extensive review of
Wilson loops see \cite{Loll} and the references therein.)
Since $T(\gamma,A)$ does not change when we
apply a gauge transformation to $A$, we may regard $T(\gamma)$ as
a function on $\A/\G$.

Ashtekar and Isham \cite{AI}
define the {\it holonomy algebra} to be the algebra of functions on
$\A/\G$ generated by piecewise smooth Wilson loops, and form the {\it
holonomy C*-algebra} as the completion of the holonomy algebra in
the $L^\infty$ norm.
One may work with continuous linear functionals on the holonomy
C*-algebra as a rigorous version of ``measures'' on $\A/\G$.
In particular, suppose that $L$ is an oriented link
in $M$ having components $(\gamma_1, \dots, \gamma_p)$, where
the $\gamma_i$ are embedded circles.  Then if $E$ is a continuous linear
functional on the holonomy C*-algebra, defining
\[      \L(L) = E(T(\gamma_1)\cdots T(\gamma_p)), \]
it is easy to see that if $E$ is invariant under the action of
$\Diff_0(M)$, $\L(L)$ is an ambient isotopy invariant of the link $L$.

However, the calculations of Polyakov \cite{Pol} and Witten
\cite{Witten} in Chern-Simons theory show
that making sense of the Chern-Simons ``measure''
\[          e^{iS(A)} {\cal D}A \]
involves further subtleties.   Polyakov gave a heuristic argument to show
that in the case of the defining representation of $G = U(1)$,
given disjoint embedded oriented circles $\gamma_1,
\gamma_2$ in $S^3$, one should be able to calculate
\[         \int_{\A/\G} T(\gamma_1,A)T(\gamma_2,A)\,
 e^{iS(A)} {\cal D}A  \]
in terms of the linking number of $\gamma_1$ and $\gamma_2$.
The argument breaks down when $\gamma_1$ and $\gamma_2$ intersect, since
one obtains a divergent integral.
If $\gamma$ is an embedded circle equipped with a framing (that is,
a section of the tangent bundle of $S^3$ over $\gamma$ that is
everywhere linearly independent from the tangent vector of $\gamma$),
one could define
\[      \int_{\A/\G} T(\gamma,A)^2\, e^{iS(A)} {\cal D}A  \]
to be the ``self-linking number'' of $\gamma$, which is an invariant of
oriented framed links.  However, since the self-linking number involves
the framing of $\gamma$, while the integral above appears not to,
this may appear rather {\it ad hoc}.

A similar problem occurs in $SU(2)$ Chern-Simons theory.
Using techniques from conformal field theory Witten argued that in
the case of the defining representation of $G = SU(2)$, one
should be able to calculate
\[          \int_{\A/\G} T(\gamma_1,A) \cdots T(\gamma_p,A)\,
 e^{iS(A)} {\cal D}A \]
in terms of the Kauffman bracket of the link in $S^3$ with
components $\gamma_i$.
Again, though, the Kauffman polynomial is an invariant of framed links.

In this paper we describe a modified holonomy
C*-algebra, and also a holonomy Banach algebra, such that
diffeomorphism-invariant continuous linear functionals on these algebras
define invariants of framed links.
Our holonomy algebras are completions of the algebra generated by a kind
of regularized Wilson loop, or ``tube.''  (Similar
regularizations appear in the physics literature; see
\cite{Ash} and the references therein.)
This clarifies the general relation between framing
and regularization in
diffeomorphism-invariant gauge theories - a
relation that is already familiar in Chern-Simons theory
\cite{AS,BGP,CGMM,GMM}.

Moreover, we show that diffeomorphism-invariant continuous linear
functionals on our holonomy algebras are classified by
the framed link invariants they define.     That is, the
map from such functionals to framed link invariants is one-to-one. (See
Theorems 2 and 4.)
One reason for working with Banach and C*-algebras
is to obtain this result, which depends on approximating
arbitrary products of tubes by products of nonintersecting tubes.
This result is a step towards finding a useful inverse of the ``loop
transform'' for diffeomorphism-invariant
gauge field theories including general relativity \cite{Loll,RS}.

I would like to thank Michel Lapidus, Geoffrey Mess, Richard Palais,
Vladimir Pestov, Michael Renardy, and Jim Stafney for their help.

\section{The Holonomy Algebra and Tube Algebra}

In all that follows, let $G$ be a compact Lie group and $\rho$ a
$k$-dimensional unitary
representation of $G$.  Let $M$ be a $n$-manifold and $P \to M$
a principal $G$-bundle over $M$.   Define $\tau \maps G \to \C$
by
\[        \tau(g) = {1\over k}\tr(\rho(g)).\]
Given a
smooth connection $A$ on $P$ and a smooth loop $\gamma \maps
S^1 \to M$, let $T(\gamma,A)$ denote the trace of the holonomy of $A$
around the loop $\gamma$, computed using the trace $\tau$.
These ``Wilson loops'' are of special interest
because they are a convenient set of gauge-invariant functions
on the space of connections.

Let us write $(t,x)$ for a point in $\torus$.  Given a point
$x \in D^{n-1}$ and a map $\gamma \maps \torus \to M$, we write $\gamma^x$
for the loop in $M$ given by
\[     \gamma^x(t) = \gamma(t,x)   .\]
We define a {\it tube} to be a pair $(\gamma, \om)$ consisting
of an embedding $\gamma \maps \torus \to
M$ and a smooth complex-valued $(n-1)$-form $\om$
compactly supported in the interior of $D^{n-1}$.
We will often use $\Gamma$ to denote a tube.
Given a tube $\Gamma = (\gamma,\om)$ and
a smooth connection $A$ on $P$, we define
\[    T(\Gamma,A) = \int_{D^{n-1}}       T(\gamma^x,A)\, \om(x)   . \]

Let $\A$ denote the space of smooth connections on $P$.  $\A$ is an
affine space with a natural topology, the $C^\infty$ topology; choosing
any point in $\A$ as an origin allows us to identify $\A$ with a locally
convex topological vector space.
This enables us to define the algebra $C^\infty(\A)$ of smooth complex
functions on $\A$.  We define the functions $T(\gamma), T(\Gamma) \in
C^\infty(A)$ for a loop $\gamma$ and a tube $\Gamma$ by:
\[   T(\gamma)(A) =  T(\gamma,A), \;\;
        T(\Gamma)(A) = T(\Gamma,A). \]

Let $\H_0$ denote the unital
 subalgebra of $C^\infty(A)$ generated by the
functions $T(\Gamma)$ for all tubes $\Gamma$.
We wish to construct a Banach algebra completion of $\H_0$ in as strong a
norm as possible, so that as many linear functionals on $\H_0$ as
possible extend to continuous linear functionals on the completion.
One might be tempted to give $\H_0$ a norm by setting
\[         \|f\| = \sup_F \|F(f)\|   \]
with supremum is taken over all homomorphisms $F$ of
$\H_0$ into a Banach algebra.  The problem is that the supremum
might not be finite.  So instead we take the supremum over a
smaller class of homomorphisms, defined using a space spanned
by equivalence classes of tubes, the so-called ``tube space,''
which we now introduce.

Let $V$ denote
the complex vector space having the set of all tubes as a basis.
Let $V_1 \subset V$ denote the subspace spanned of finite linear
combinations
\[         \sum_i c_i (\gamma,\om_i)  \]
such that
\[        \sum_i c_i \om_i = 0 .\]
Let $V_2 \subset V$ denote the subspace spanned by elements of the
form
\[        (\gamma_1,\om_1) - (\gamma_2,\om_2)  \]
where $(\gamma_1,\om_1)$ and $(\gamma_2,\om_2)$ are {\it essentially
equivalent}, that is, there are embedded discs $D_1,D_2 \subseteq D^{n-1}$
with $\supp\,\om_i \subseteq D_i$, and a diffeomorphism $\alpha
\maps D_1 \to D_2$ such that
\[            \om_1 = \alpha^\ast \om_2  \]
and such that
for all $x \in D_1$, $\gamma_1^x$ is equal to $\gamma_2^{\alpha(x)}$
up to an orientation-preserving reparametrization of $S^1$.

Define the vector space $\Tubes_0$ by
\[             \Tubes_0 =  V/(V_1 + V_2)  .\]
We give $\Tubes_0$ a norm as follows:
\[        \|v\| = \inf \{ \,\sum_i \|c_i\om_i\| \colon\;
v = \sum_i c_i [(\gamma_i, \om_i)] \, \} .\]
Here we use an $L^1$-type norm on $(n-1)$-forms on $D^{n-1}$ given by
\[            \|\om\| = \int_{D^{n-1}} |f(x)| \, dx  \]
where $dx$ is the standard volume form on $D^{n-1}$ and
$F$ is the function such that
$                \om = f dx $.
Let the {\it tube space} $\Tubes$ denote the
Banach space completion of $\Tubes_0$.

There is a linear map $T \maps V \to \H_0$
given by
\[          T(\Gamma)(A) = T(\Gamma,A)  .\]
Moreover:

\begin{lemma}\DOT \label{lem1}
 The map $T \maps V \to \H_0$ vanishes on $V_1 +
V_2$. \end{lemma}

Proof -
Note that if $\sum c_i (\gamma,\om_i) \in V_1$, then
\[         T(\sum c_i (\gamma,\om_i))(A) =
\sum\int_{D^2} T(\gamma^x,A)  c_i \om_i(x)
=     0  , \]
so $T$ vanishes on $V_1$.
Also, if $\Gamma_1 =
(\gamma_1,f_1)$ and $\Gamma_2 = (\gamma_2,f_2)$ are essentially
equivalent tubes, with $D_1,D_2,\alpha$ as above, then
\ba  T(\Gamma_1)(A) &=&
   \int_{D_1}       T(\gamma_1^x,A)\, \om_1(x) \nonumber\cr
&=&\int_{D_1} T(\gamma_2^{\alpha(x)},A)\, (\alpha^\ast\om_2)(x)
\nonumber\cr
&=&  \int_{D_2} T(\gamma_2^x,A) f_2(x) \nonumber\cr
&=&  T(\Gamma_2,A)  \nonumber\ea
Thus $T$ vanishes on $V_2$.
\qed

It follows that $T$ factors through $\Tubes_0$;
that is, we may define a linear map, which
we also call $T$, from $\Tubes_0$ to $C^\infty(\A)$ by
\[        T([\Gamma])(A) = T(\Gamma)(A)   .\]
Henceforth, given a tube $\Gamma = (\gamma,\om)$, we will often abuse
notation and write simply $\Gamma$ or
$(\gamma,\om)$ for its equivalence class in $\Tubes_0$.

Now we return to the problem of putting a norm on $\H_0$.
Let $\Lambda$ denote the class of all homomorphisms $F \maps
\H_0 \to A$, where $A$ is a Banach algebra, such that
\[         \|F(T(v))\| \le \|v\|  \]
for all $v\in \Tubes_0$.   Give $\H_0$ a norm as follows:
\[           \|f\| = \sup_{F\in \Lambda} \|F(f)\|  . \]
Note that $\|f\|$ is finite for all $f \in \H_0$.  Also, note that
$\iota \in \Lambda$, where
$\iota$ is the inclusion of $\H_0$ in the C*-algebra of
bounded continuous functions on $\A$,
and that $\iota(f) \ne 0$ for all nonzero $f \in
\H_0$.  Thus $\|f\|$ is nonzero for all nonzero $f \in \H_0$, so
$\|\cdot\|$ is indeed a norm on $\H_0$.  The completion of
$\H_0$ in this norm is a Banach algebra, the
{\it holonomy Banach algebra}, $\H_b$, associated
to the bundle $P$ and the representation $\rho$.

We will always use $\|\cdot\|$ to denote the above norm on $\H_0$, and
use $\|\cdot\|_\infty$ to denote the $L^\infty$ norm on $\H_0$ regarded
as a subalgebra of $B(\A)$.  By the above,
\[            \|f\|_\infty \le \|f\| \]
for all $f \in \H_0$, so $\H_b$ may be regarded as a subalgebra of
the bounded continuous functions on $\A$.   Thus $\H_b$ is a commutative
semisimple Banach algebra with unit.
In Section 5, we will discuss a
holonomy C*-algebra, $\H_c$.   This is just the completion of $\H_0$
in the $L^\infty$ norm.
The diffeomorphism-invariant continuous linear
functionals on both $\H_b$ and $\H_c$ are classified by invariants of
framed links in $M$.  Each algebra
has its own technical advantages.  From the viewpoint of mathematical
physics, it is preferable to work with a C*-algebra as an
``algebra of observables'' for gauge field theories, since
C*-algebras have a well-understood representation theory that is closely
tied to Hilbert space theory.
On the other hand, the Banach algebra approach is in a
sense more conservative: $\H_b$ is a dense subalgebra of
 $\H_c$, so every continuous linear
functional on $\H_c$ defines one on $\H_b$.

To relate the holonomy Banach algebra to link invariants it is
convenient to introduce an auxiliary object, the ``tube Banach algebra,''
depending only on the base manifold $M$.
First, note that by definition of the norm on $\H_b$,
\[           \|Tv\| \le \|v\|  \]
for all $v \in \Tubes_0$.  It follows that $T$
extends uniquely to a continuous linear map
from $\Tubes$ to $\H_b$, which will also call $T$.

\begin{question}\DOT  What is the kernel of the map $T \maps \Tubes_0
\to \H_0$?
\end{question}

\begin{question}\DOT  What is the kernel of the map $T \maps \Tubes
\to \H_b$?
\end{question}

The map $T \maps \Tubes \to \H_b$ extends uniquely
to an algebra homomorphism, which we again call $T$, from the symmetric
algebra $S\Tubes$ to $\H_b$.  We now construct the ``tube Banach algebra''
and further extend $T$ to a homomorphism from this algebra
to $\H_b$.  For this we use the following
general result about completing the symmetric algebra over a Banach
space.   By a {\it contraction} we mean a linear
map $f \maps X \to Y$ between Banach spaces satisfying $\|f(x)\| \le
\|x\|$. Also, in all that follows Banach algebras will be assumed to
have a multiplicative unit, and homomorphisms between them
will be assumed to preserve the unit.

\begin{lemma}  \DOT \label{lem2}
Let $X$ be a Banach space.   Then
up to isometric isomorphism there is a unique
commutative Banach algebra $S_bX$ equipped with a
contraction $\iota \maps X \to S_bX$ such that:
given any contraction $F \maps X \to A$ where
$A$ is a commutative Banach algebra $A$, there is a unique
homomorphism $\widetilde F \maps S_bX \to A$ that is a contraction
and such that $F = \widetilde F \iota$.
Moreover, the space $\iota(X) \subseteq X$ generates a dense subalgebra
of $S_bX$ naturally isomorphic to the symmetric algebra $SX$.
\end{lemma}

Proof -   Uniqueness follows from the universal
property of $S_bX$. For existence, we construct $S_bX$ as
follows.
Let $\Lambda$ be the class of homomorphisms $F \maps SX
\to A$, $A$ a commutative Banach algebra, such that
\[          \|F(x)\| \le \|x\|  \]
for all $x \in X$.  We define a norm on $SX$ as follows:
\[      \|f\| = \sup_{F \in \Lambda} \|F(f)\|  \]
for all $f \in SX$.
To show that this norm is well-defined it suffices to show that the
supremum is finite, and nonzero for every nonzero $f$.
Finiteness follows from the fact that  $\|F(x)\| \le \|x\|$ for
every $x \in X$, $F\in \Lambda$.  To prove that $\|f\| > 0$ for all
nonzero $f \in SX$, it suffices to find $F\in \Lambda$ that is one-to-one.
We defer this to the proof of Lemma \ref{lem3'} below.

Now let $S_bX$ be the completion of $SX$ in the norm
we have constructed.  Then $S_bX$ is a commutative Banach algebra,
and the natural embedding $\iota \maps X \to S_bX$ has
the desired universal property.  It also follows from the
construction that $\iota(X)$ generates a dense subalgebra of $S_bX$
isomorphic to $SX$.
   \qed

It follows from Lemma \ref{lem2} that the homomorphism $T \maps
S\Tubes \to \H_b$ extends uniquely to a continuous homomorphism
from $S_b\Tubes$ to $\H_b$.  We again call this $T$.

\section{Invariant States on the Holonomy Algebra}

Let $\Aut(P)$ denote the group of smooth bundle automorphisms of $P$,
not necessarily base-preserving.  Let
 $\Aut_0(P)$ denote the
identity component of $\Aut(P)$.  The group
 $\Aut(P)$ acts as diffeomorphisms of
$\A$, hence as isomorphisms of the algebra $\H_b$.  Note
that we have an exact sequence
\[      1 \to \G \to \Aut(P) \to \Diff(M) \to 1 \]
Since the action of $\G$ on $\H$ is trivial, $\Diff(M)$ acts as
automorphisms of $\H_b$ as well.  Our goal is to
classify generalized ``measures'' on $\A/\G$
 that are invariant under the action of $\Diff_0(M)$.  We will require of
such ``measures'' $\mu$ only that the integral
\[           \int_{\A/\G} f(A)\, d\mu(A)  \]
make sense for all $f \in \H_b$ (where we identify $\H_b$ with
an algebra of functions on $\A/\G$) and depend linearly and
continuously on $f$.  More precisely,
we seek to classify $\mu \in \H_b^\ast$ that satisfy
\[           \mu(f) = \mu(gf)  \]
for all $f \in \H_b$, $g \in \Diff_0(M)$.  We will show that there is a
natural one-to-one mapping from such functionals to complex-valued
ambient isotopy invariants of framed oriented links in $M$.
Unfortunately, the range of this mapping is unknown even in the simplest
cases, so the classification program is far from complete.

The strategy is as follows.  To begin with, note that
the group $\Diff(M)$ acts as isometries of $\Tubes$, hence as
automorphisms of the tube algebra
$S\Tubes$.  Also, the homomorphism
$   T \maps S_b\Tubes \to \H_b $
satisfies
\[              T(ga) = g T(a) \]
for all $g \in \Diff(M)$, $a \in S_b\Tubes$.
It follows that every $\Diff_0(M)$-invariant
functional $\mu \in \H_b^\ast$ gives rise to a $\Diff_0(M)$-invariant
linear functional $\mu\circ T \in S_b\Tubes$.
Moreover, since $T$ has dense range, its adjoint
\[              T^\ast \maps \H_b^\ast \to (S_b\Tubes)^\ast  \]
is one-to-one.  In what
follows we shall show that there is a one-to-one map from the space of
$\Diff_0(M)$-invariant elements of
$(S_b\Tubes)^\ast$ to the space of complex-valued ambient isotopy
invariants of oriented framed links in $M$.
As a corollary, then, we obtain a one-to-one map from $\H_b^\ast$
to this space of link invariants.

Given a set of tubes $\{\Gamma_i\}$, where
$\Gamma_i = (\gamma_i, \om_i)$, we say that the $\Gamma_i$ are {\it
nonintersecting} if the embeddings
$\gamma_i \maps \torus \to M$ have disjoint ranges.  Let
$\om= f dx$ be a fixed but arbitrary smooth
$(n-1)$-form supported in the interior of $D^{n-1}$ with $f \ge 0$ and
\[        \int_{D^{n-1}} \om = 1 .\]
We say that the set of tubes $\lbrace\Gamma_i\rbrace$ is {\it
normalized} if $\om_i = \om$ for all $i$.
The following lemma is
the basis for the relation between diffeomorphism-invariant elements
of $(S_b\Tubes)^\ast$ and link invariants.

\begin{lemma} \label{lem6} \DOT  Suppose
$M$ is a manifold of dimension $\ge 3$.    Then
 elements of the form $\Gamma_1 \cdots
\Gamma_p$, where the $\Gamma_i = (\gamma_i,\om)$
are nonintersecting normalized tubes in $M$, span a subspace
of the tube algebra $S_b\Tubes$ that is dense in the norm topology.
 \end{lemma}

Proof - By Lemma \ref{lem2}, products of tubes  $\Gamma_1 \cdots \Gamma_p$
span a dense space of  $S_b\Tubes$.   Since every smooth
$(n-1)$-form compactly supported in the interior
of $D^{n-1}$ can be approximated in the $L^1$ norm by a linear
combination of $(n-1)$-forms of the form $\alpha^\ast \om$, where
$\alpha$ is a diffeomorphism of $D^{n-1}$, it follows that
linear combinations of normalized tubes are dense in the space of
tubes.  Thus it suffices to show that each product of
tubes $\Gamma_1 \cdots
\Gamma_p$ may be approximated in the norm topology on $\H$ by finite
linear combinations
\[          \sum_\ell \Gamma_1^\ell \cdots \Gamma_p^\ell  \]
where for each $\ell$, the tubes $\Gamma_i^\ell$ are nonintersecting.
We use the following fact from differential topology:

\begin{lemma}\label{lem7}\DOT  Let $\gamma_1, \dots, \gamma_p$ be
embeddings of $\torus$ in $M$.  Let $U \subseteq (D^{n-1})^p$ be the set
of $p$-tuples $(x_1, \dots, x_p)$ such that the ranges of the
loops $(\gamma_i)^{x_i} \maps S^1 \to M$ for $i = 1,\dots, p$ are
disjoint.  Then $U$ is an open dense set in $(D^{n-1})^p$ if $n \ge 3$.
\end{lemma}

Proof - Note that
\[
 \gamma_1\times\cdots\times\gamma_p \maps (\torus)^p \to M^p  \]
is an embedding.
Let $D$ be the diagonal in $M^2$, and suppose $i\ne j$.
  Then
\[B_{ij} =  (\gamma_i \times\gamma_j)^{-1} D \]
 is an $n$-dimensional submanifold of $(\torus)^2$.
Let $\pi \maps \torus \to D^{n-1}$ denote projection onto the second
factor.
It follows that
\[      (\pi\times \pi)B_{ij} \subset (D^{n-1})^2 \]
has an open dense complement
$C_{ij}$, since the dimension of $(D^{n-1})^2$ is greater than $n$.
The set $C_{ij}$ is precisely the set of pairs $(x,y)$ such that
the ranges of $\gamma^x_i$ and $\gamma^y_j$ are disjoint.
Letting $p_{ij} \maps (D^{n-1})^p \to (D^{n-1})^2$ be given
by $p_{ij}(x_1, \dots, x_n) = (x_i,x_j)$, it follows that
$p_{ij}^{-1} C_{ij}$ is open and dense in $(D^{n-1})^p$.  Since
 $U$ is the intersection of the sets
 $p_{ij}^{-1}C_{ij}$ it is open and dense.
\qed

Now let $\Gamma_i = (\gamma_i, \om_i)$, $1 \le i \le p$, be a collection
of tubes in $M$.  Let $K \subset (D^{n-1})^p$ be the intersection of
$\supp\, \om_1 \times \cdots \times \supp\, \om_p$ with the complement
of the set $U$ given in Lemma \ref{lem7}.  $K$ is compact and nowhere
dense.   Choose $\eps > 0$.  Then we may cover $\supp\, \om_i \subset
D^{n-1}$ with embedded discs $D_{ij}$, $1 \le j \le k$, and write
\[         \om_i = \sum_{j = 1}^k \om_{ij} \]
where $\om_{ij}$ is a smooth form supported in the interior of
$D_{ij}$, in such a way that the sum of
\[           \|\om_{1j_1}\| \cdots  \|\om_{pj_p}\|   \]
over all $(j_1, \dots , j_p)$ with
\[          D_{1j_1} \times \cdots \times D_{pj_p} \cap K \ne \emptyset \]
is $\le \eps$.   Let $J$ denote the set of such $p$-tuples
$(j_1, \dots , j_p)$, and let $\Gamma_{ij} = (\Gamma_i, f_{ij})$.
Then
\[        \|\Gamma_{1j_1} \cdots \Gamma_{pj_p}\|
\le  \|f_{1j_1}\| \cdots  \|f_{pj_p}\| ,\]
so writing
\[  \Gamma_1 \cdots \Gamma_p
=  \sum_{(j_1, \cdots, j_p) \in J} \Gamma_{1j_1} \cdots \Gamma_{pj_p}
+ \sum_{(j_1, \cdots, j_p) \notin J} \Gamma_{1j_1} \cdots
\Gamma_{pj_p},\]
it follows that
\[    \|  \Gamma_1 \cdots \Gamma_p - \sum_{(j_1, \cdots, j_p) \notin J}
\Gamma_{1j_1} \cdots \Gamma_{pj_p} \| \le \eps.  \]
Given $(j_1, \dots, j_p) \notin J$, there are nonintersecting
 tubes $\Gamma'_{1j_1},
\dots ,\Gamma'_{pj_p}$ such that $\Gamma'_{ij_\ell}$ is essentially
equivalent to $\Gamma_{ij_\ell}$.  Moreover,
\[    \|  \Gamma_1 \cdots \Gamma_p - \sum_{(j_1, \cdots, j_p) \notin J}
\Gamma'_{1j_1} \cdots \Gamma'_{pj_p} \| \le \eps.  \]
as desired.  \qed

It follows that in dimension $\ge 3$ any continuous linear functional $\mu
\maps S_b\Tubes \to \C$ is determined by its values on products
 of nonintersecting normalized tubes.  Next, note that nonintersecting
tubes determine framed oriented links in $M$ as follows.
Given a tube $(\gamma,\om)$ in $M$, the loop $\gamma^0 \maps S^1 \to M$
is an embedded circle, where $0 \in D^{n-1}$ is the origin.
This embedded circle acquires an orientation from the
standard orientation on $S^1$, and its conormal bundle has a frame given
by
\[     \{(d\gamma)_{(t,0)}(0,e_j)\}_{j=1}^{n-1}   \]
where $e_j$ are the standard basis of the tangent space at the origin of
$D^{n-1}$.   If $\Gamma_1, \dots, \Gamma_p$
are nonintersecting tubes, the
embedded circles $\gamma_j^0$ are disjoint, so we obtain a framed
oriented link with $p$ components.  Let us write this link as
$L(\Gamma_1, \dots, \Gamma_p)$.

Recall that two oriented framed links $L,L'$
in $M$ are said to be ambient isotopic
if there exists $f \in \Diff_0(M)$ taking $L$ to $L'$,
including orientation and framing.  We will write this simply as
$fL = L'$.

\begin{lemma}\label{lem8} \DOT Let $\Gamma_1, \dots, \Gamma_p$ be
nonintersecting normalized
tubes in the manifold $M$, and let $\Gamma'_1, \dots, \Gamma'_p$ also be
nonintersecting normalized tubes in $M$.  Then the framed oriented
links $L = L(\Gamma_1, \dots, \Gamma_p)$ and
$L' = L(\Gamma'_1, \dots, \Gamma'_p)$ are ambient isotopic if and
only if, possibly after some reordering of the indices of the
$\gamma_i$, there exists $f \in \Diff_0(M)$ such that
\[         f\Gamma_i = \Gamma'_i  \]
for all $i$.
  \end{lemma}

Proof - It is clear that if there exists $f \in \Diff_0(M)$ with
$f \Gamma_i = \Gamma'_i$ for all $i$ (up to a reordering of indices),
then $L = L(\Gamma_1, \dots,
\Gamma_p)$ and $L' = L(\Gamma'_1, \dots, \Gamma'_p)$ are ambient
isotopic as framed oriented links.

For the converse, suppose the framed oriented
links  $L$ and $L'$ are ambient isotopic.  Let $B$ denote the disjoint
union of $p$ copies of $S^1$, $D$ denote the disjoint union of $p$
copies of $\torus$, and $E$ denote the disjoint union of $p$ copies of
$S^1 \times \R^{n-1}$.  We regard $E$ as a vector bundle over $B$ and
$D$ as the unit disc bundle.  Let $g, g' \maps D
\to M$ denote the embeddings defined by the nonintersecting
 tubes $\{\Gamma_i\}$ and
$\{\Gamma'_i\}$, respectively.  Then, possibly after a reordering of
indices, there exists $f \in \Diff_0(M)$ with
\[        f\circ g|_B = g'|_B \]
and
\[       d(f\circ g)|_B = dg'|_B .\]
Let us write simply $g$ for $f \circ g$.  Then we need to show that
there is a diffeomorphism $F \in \Diff_0(M)$ such that $F \circ g = g'$.
By the isotopy extension theorem \cite{Palais}
it suffices to find an isotopy $g_t
\maps D \to M$ such that $g_0 = g$ and $g_1 = g'$.  Using tubular
neighborhoods of $gD, g'D$ we can extend $g,g'$ to embeddings
$G,G' \maps E \to M$, and it clearly suffices to find an isotopy $G_t
\maps E \to M$ with $G_0 = G$ and $G_1 = G'$.   Since $G$ and $G'$ agree
on $B$, as do $dG$ and $dG'$, the construction used in the proof of the
uniqueness of tubular neighborhoods (in \cite{Hi}, for example) gives
the desired isotopy $G_t$.  \qed

We thus obtain:

\begin{theorem} \label{thm1}\DOT  Let $S_b\Tubes$ be the
tube Banach algebra of $M$, and suppose that $\dim M \ge 3$.
Suppose that $\mu \in (S_b\Tubes)^\ast$ is invariant
under the action of $\Diff_0(M)$.  Then there is a complex-valued
ambient isotopy
invariant $\L_\mu$ of framed oriented links in $M$ given by
\[           \L_\mu(L) = \mu(\Gamma_1 \cdots\Gamma_p) \]
where $\Gamma_1 ,   \cdots , \Gamma_p$ are any choice of normalized
nonintersecting tubes for which $L = L(\Gamma_1, \dots, \Gamma_p)$.
Moreover, the map $\mu \mapsto \L_\mu$ is one-to-one.   \end{theorem}

Proof - By Lemma \ref{lem8}, $\L_\mu$ is well-defined and an ambient
isotopy invariant of framed oriented links in $M$.   If
$\L_\mu = 0$, then $\mu$ vanishes on all products of normalized
nonintersecting tubes, so by Lemma \ref{lem7}, $\mu = 0$. \qed

We remark that it is easy to show that $L_\mu$ is independent of the
choice of $(n-1)$-form $\om$ used in the definition of normalized tube.

\begin{theorem} \label{thm2}  \DOT  Let $P$ be a principal $G$-bundle
over a manifold $M$ of dimension $\ge 3$,
 and $\rho$ a finite-dimensional unitary
representation of $G$.  Let $\H_b$ be the holonomy Banach
algebra associated to this data.
The map from $\Diff_0(M)$-invariant elements of
$\H_b^\ast$ to complex invariants of framed oriented links in $M$
given by
\[           \mu \mapsto \L_{T^\ast\mu}  \]
is one-to-one.  \end{theorem}

Proof - This follows immediately from the theorem and the fact that
$T^\ast\maps \H_b^\ast \to (S_b\Tubes)^\ast$ is one-to-one.  \qed

We are left with two questions:

\begin{question}\DOT \label{q1} What is the range of the map $\mu \mapsto
\L_\mu$ given in Theorem \ref{thm1}?  \end{question}

\begin{question}\DOT \label{q2}
 What is the range of the map $\mu \mapsto \L_{T^\ast \mu}$ given in
Theorem \ref{thm2}?  \end{question}

In particular, it would be interesting to know whether the link invariants
given by Theorem \ref{thm2} include those given by Chern-Simons theory
when $M$ is compact.
It would also be interesting to know whether the link invariants given
by this theorem form a complete set of invariants as $P$ ranges
over all bundles with compact gauge group and $\rho$ ranges over
all finite-dimensional unitary representations.   These questions are
related, via recent conjectures concerning Vassiliev invariants
\cite{Bar-Natan,Birman,Vassiliev}.

We defer a serious study of the above questions to future work,
contenting ourselves here with two remarks.  First, recall that
the theory of links is rather trivial in
dimension $> 3$.   More precisely, the ambient isotopy
classes of framed oriented knots in an oriented manifold of dimension
$> 3$ are in one-to-one correspondence (although not naturally) with
elements of $\pi_1(M) \times \Z_2$
\cite{R}.  Here $\pi_1(M)$ records the homotopy class of
the knot, while $\Z_2$ records the framing, essentially because
$\pi_1(SO(n-1)) \simeq \Z_2$.   Similarly, ambient isotopy classes of
framed oriented links correspond to unordered $n$-tuples of elements in
$\pi_1(M) \times \Z_2$.  It thus appears
that diffeomorphism-invariant ``measures'' on
the space of connections modulo gauge transformations are more easily
classified in dimension $>3 $ than in dimension $3$, where one hopes that
Chern-Simons theory provides
interesting examples of such measures.

Second, there is a simple example of Theorem \ref{thm2}
involving the moduli space of flat connections.  This
has been treated already in the $SU(2)$ case by Ashtekar and Isham
\cite{AI}, working with their holonomy
C*-algebra.  Let $M$ be any manifold, $G$ a compact Lie group,
and $P \mapsto M$ a $G$-bundle over $M$.  Let $\A_0$ denote the space of
smooth flat connections on $P$.  Then $\A_0/\G$ is a subspace of
$\A/\G$ that may be identified with $\Hom(\pi_1(M),G)/G$, where $G$ acts
by conjugation.  A finite Borel
measure $d\mu$ on $\A_0/\G$ defines a linear functional
$\mu_0 \maps \H_0 \to \C$ as follows:
\[         \mu_0(f) = \int_{\A_0/\G} f(A)\, d\mu(A)  .\]
Since
\[           |\mu_0(f)| \le c\|f\|_\infty \]
for some $c > 0$, where
\[           \|f\|_\infty = \sup_{A \in \A} |f(A)|  ,\]
and since this $L^\infty$ norm is weaker than the norm in which we
complete $\H_0$ to obtain $\H_b$, it follows that $\mu_0$ extends
uniquely to a continuous linear functional $\mu \maps \H_b \to \C$.
Since the group $\Diff_0(M)$ acting on $\A/\G$ fixes $\A_0/\G$, the
functional $\mu$ is $\Diff_0(M)$-invariant.  Thus by Theorem \ref{thm2}
the functional $\mu$
determines a link invariant $\L_{T^\ast \mu}$.  This link invariant only
depends on the homotopy classes of the components of the link.

\section{C*-algebraic considerations}

In this section we describe C*-algebraic analogs of the results on
holonomy Banach algebras.
Recall that $\H_0$ denotes the unital
subalgebra of $C^\infty(A)$ generated by the
functions $T(\Gamma)$ for all tubes $\Gamma$.  We claim that
$\H_0$ is a sub-$\ast$-algebra of $C^\infty(A)$, the latter being
a $\ast$-algebra under pointwise complex conjugation.
Given an embedded solid torus $\gamma \maps \torus \to M$, define
the embedded solid torus $\overline\gamma \maps \torus \to M$ by
\[      \overline\gamma(t,x) = \gamma(- t,x)  ,\]
where we identify $S^1$ with $\R/\Z$.  Given a tube $\Gamma =
(\gamma,f)$, let
\[          \overline\Gamma = (\overline\gamma,\overline \om)  \]
where $\overline \om$ denotes the pointwise complex conjugate of the
differential form $\om$.
Then we have
\ba   \overline{  T(\Gamma, A)} &=&  \int_{D^{n-1}}
\overline{T(\gamma^x,A)}\,\, \overline \om(x)  \nonumber\cr
&=&      \int_{D^{n-1}} T(\overline \gamma^x,A)\, \overline
\om(x)\nonumber\cr &=&     T(\overline\Gamma,A)  ,\nonumber\ea
where we use the fact that the complex conjugate of
${\tr(g)}$ is $\tr(g^{-1})$ for
$g$ unitary.   It follows that $\H_0$
is a sub-$\ast$-algebra of $C^\infty(\A)$.

Note that the
functions $T(\Gamma,\cdot)$ are bounded, so the $L^\infty$ norm on
$\H_0$, given by
\[           \|f\|_\infty = \sup_{A \in \A} |f(A)|, \]
is well-defined, and
the completion of $\H_0$ in this norm is a C*-algebra.
We call this C*-algebra the {\it holonomy C*-algebra} associated
to the bundle $P$ and the representation $\rho$, and denote it by
$\H_c$.   The holonomy algebra may
be regarded as the closure of $\H_0$ in the C*-algebra of
bounded continuous functions on $\A$.

It is worth noting the differences between the above holonomy C*-algebra
and that defined by Ashtekar and Isham \cite{AI}
in the special case
where $G = SU(2)$ and $\rho$ is the defining (2-dimensional)
representation.  In this case,
\[           T(\overline\gamma, A) = T(\gamma,A) \]
for all piecewise smooth loops $\gamma$ and connections $A$,
where $\overline\gamma$ is $\gamma$ with its orientation reversed, so
one does not need to reverse orientations to define the $\ast$-algebra
structure of the holonomy algebra.
However,
using orientation-reversed loops one could define
holonomy C*-algebras of the Ashtekar-Isham type quite generally.
Namely, the algebra generated by functions of
the form $T(\gamma)$, where $\gamma$ is a piecewise smooth loop,
is a sub-$\ast$-algebra of the bounded continuous functions on $\A$,
and its completion in the $L^\infty$ norm is a
C*-algebra.  It is easy to see that any $\Diff_0(M)$-invariant
continuous linear functional on this C*-algebra determines a link
invariant.  However, such link invariants will not depend on a framing.
Thus we do not expect Chern-Simons theory to define continous linear
functionals on this type of holonomy C*-algebra.  Moreover, one does not
expect the map from $\Diff_0(M)$-invariant
continuous linear functionals to link invariants to be one-to-one in
this case, because there is no analog of Lemma \ref{lem6}.

In what follows we define a ``tube C*-algebra'' and extend the results
of Sections 2 and 3 to the C*-algebraic setting.
The algebra $S\Tubes$ has a natural
$\ast$-algebra structure making $T\maps S\Tubes \to \H_0$ a
$\ast$-homomorphism.  To see this, note the following:

\begin{lemma} \label{lem1'} \DOT Let $K \maps V \to V$ be the
conjugate-linear map given by $K(\Gamma) = \overline \Gamma$ for all
tubes $\Gamma$. Then $KV_1 \subseteq V_1$ and  $KV_2 \subseteq V_2$.
\end{lemma}

Proof -  For $V_1$, given a linear combination of
tubes $v = \sum c_i (\gamma,\om_i)$ such that
$\sum_i c_i f_i = 0$,
note that  $\sum_i \overline c_i \overline \om_i = 0$ implies
$Kv \in V_1$.
For $V_2$, note that if $\Gamma$ and $\Gamma'$ are essentially
equivalent tubes, so are $\overline\Gamma$ and $\overline\Gamma'$.  \qed

We thus obtain a conjugate-linear map $K \maps \Tubes_0 \to \Tubes_0$.

\begin{lemma} \label{lem2'} \DOT  The map $K \maps \Tubes_0 \to \Tubes_0$
is norm-preserving.  \end{lemma}

\begin{lemma} \label{lem2'} \DOT  The map $K \maps \Tubes_0 \to \Tubes_0$
is norm-preserving.  \end{lemma}

Proof - Given $v\in \Tubes_0$, for any $\eps > 0$ we may
write
\[            v = \sum c_i (\gamma_i,\om_i) \]
with
\[         \sum_i \|c_i \om_i\|  \le \|v\| + \eps  .\]
We then have
\[           Kv = \sum \overline c_i (\overline \gamma_i,\overline \om_i)
\]
hence
\[       \|Kv \| \le \sum_i \|\overline c_i \overline \om_i\| \le
\|v\| + \eps .\]
Since $\eps$ was arbitrary it follows that $\|Kv\| \le \|v\|$.  Since
$K^2 = 1$, $\|Kv\| = \|v\|$.
\hbox{\hskip 30 em}\qed

It follows that $K$ extends uniquely to a continuous (in fact
norm-preserving) conjugate-linear map $K \maps \Tubes \to \Tubes$.
Thus the symmetric algebra $S\Tubes$ becomes a $\ast$-algebra in a
unique manner such that
\[                  v^\ast = Kv  \]
for $v \in \Tubes$.  This implies that $T \maps S\Tubes \to \H_c$
is a $\ast$-homomorphism.

Next we use a general construction.
Given a Banach space $X$, we define a {\it conjugation}
$\kappa \maps X \to X$ to be a continuous conjugate-linear map with
$\kappa^2 = 1$.  We define a {\it $\ast$-contraction} $f \maps X
\to A$ from a Banach space $X$ with conjugation $\kappa$ to a Banach
$\ast$-algebra $A$
to be a linear map with $f(\kappa x) = f(x)^\ast$ and $\|f(x)\| \le
\|x\|$ for all $x \in X$.  Then we have:

\begin{lemma}  \DOT \label{lem3'}
Let $X$ be a Banach space with conjugation $\kappa$.  Then
up to isomorphism there is a unique
commutative C*-algebra $S_cX$ equipped with a $\ast$-contraction
$\iota \maps X \to S_cX$ such that
given any commutative C*-algebra $A$ and $\ast$-contraction
$F \maps X \to A$, there exists a unique
homomorphism $\widetilde F \maps S_cX \to A$ for which the
$F = \widetilde F \iota$.
Moreover
the subspace $\iota(X) \subseteq X$ generates a dense subalgebra
of $S_cX$ naturally isomorphic to the symmetric algebra $SX$.
\end{lemma}

Proof - Uniqueness follows from the universal
property.     For existence, we construct $S_cX$ as
follows.  First, make $SX$ into a $\ast$-algebra in the unique manner
such that $x^\ast = \kappa(x)$ for all $x \in X$.
Let $\Lambda$ be the class of $\ast$-homomorphisms $F \maps SX
\to A$, $A$ a commutative C*-algebra, such that
\[          \|F(x)\| \le \|x\|  \]
for all $x \in X$.  We define a norm on $SX$ as follows:
\[      \|F\| = \sup_{F \in \Lambda} \|F(f)\|  \]
for all $f \in SX$.
To show that this norm is well-defined it suffices to show that the
supremum is finite and nonzero for every nonzero $f$.
Finiteness follows from the fact that  $\|F(x)\| \le \|x\|$ for
every $x \in X$, $F\in \Lambda$.  To prove that $\|f\| > 0$ for all
nonzero $f \in SX$, we construct a faithful representation
$F\in \Lambda$ as follows.

Let $X_r \subseteq X$, the ``real part'' of $X$, be the closed real
subspace
\[          X_r = \{x \in X\colon \; \kappa(x) = x\}  .\]
Given $x \in X$, define $\Re(x) = (x + \kappa(x))/2 \in X_r$ and
$\Im(x) =  (x - \kappa(x))/2i \in X_r$.  Elements of
the real dual $X_r^\ast$ may be identified with elements of
$X^\ast$  as follows:
\[            \ell(x) = \ell(\Re(x)) + i\ell(\Im(x)) \]
for $\ell \in X_r^\ast$ and $x \in X$.   This allows us to identify
$X_r^\ast$ with the real subspace of $X^\ast$ consisting of functionals
 $\ell$ such that $\ell(x)$ is real for all $x \in X_r$.

Let $B$ denote the closed unit ball in $X_r^\ast$
with respect to the $X^\ast$ norm, and
give $B$ the topology in which $\ell_\alpha \to \ell$ if
$\ell_\alpha(x) \to \ell(x)$ for all $x \in X$.
Then there is a unique homomorphism
$F\maps SX \to C(B)$ with
\[ F(x)(\ell) = \ell(x) \]
for all $x \in X, \ell \in B$.    Note that $F$
is a $\ast$-homomorphism because
\ba          F(x^\ast)(\ell)
&=&     \ell(\kappa(x)) \nonumber\cr
&=&     \ell(\Re(\kappa(x))) + i\ell(\Im(\kappa(x))) \nonumber\cr
&=&     \ell(\Re(x)) - i\ell(\Im(x)) \nonumber\cr
&=&     \overline { F(x)(\ell)}  \nonumber\ea
Note also that $F \in \Lambda$ since
\[       \|F(x)\| = \sup_{\ell \in B} |\ell(x)|     \le \|x\|.  \]

We claim that $F$ is faithful.
Suppose that $f \in SX$ is nonzero.  Then $f$ is a polynomial in
elements of some finite-dimensional subspace $V \subseteq X$.
Let $V_r = V \cap X_r$, and identify $V_r^\ast$ as a real subspace of
$V^\ast$ just as we identified $X_r^\ast$ with a real subspace of
$X^\ast$.  Then since we are in a finite-dimensional
situation we can find $\ell \in V_r^\ast$ with $\|\ell\| \le 1$
and $f(\ell) \ne 0$.  By the Hahn-Banach theorem we can extend
$\ell$ to an element in $X_r^\ast \cap B$,
which we again call $\ell$.   It follows that $F(f)(\ell)
\ne 0$, so $F(f) \ne 0$.

Now let $S_cX$ be the completion of $SX$ in the norm
we have constructed.  Then $S_cX$ is a C*-algebra,
and the universal property follows immediately.  It also follows from the
construction that $\iota(X)$ generates a subalgebra of $S_cX$
isomorphic to $SX$.  \hbox{\hskip 30em}\qed

We call $S_c\Tubes$ the {\it tube C*-algebra}.  Note
that it depends only on the base manifold $M$.  Also, since any
isomorphism of commutative C*-algebras is an isometric
$\ast$-isomorphism, it is unique up to isometric $\ast$-isomorphism.

\begin{lemma} \label{lem4'}\DOT The map $T\maps
\Tubes_0 \to \H_0$ satisfies $\|Tv\|_\infty \le \|v\|$ for all $x \in
\Tubes_0$.   \end{lemma}

Proof -
Given $v\in \Tubes_0$, for any $\eps > 0$ we may
write
\[            v = \sum c_i (\gamma_i,\om_i) \]
in such a way that
\[         \sum_i \|c_i \om_i\| \le \|v\| + \eps  .\]
Then
\ba       \|Tv\|_\infty &=&
 \|\sum c_i T(\gamma_i,\om_i) \|_\infty \nonumber\cr
 &=&
\sup_{A \in \A}  | \sum_i c_i \int_{D^{n-1}}  T(\gamma_i^x,A)\, \om_i(x)|
\nonumber\cr
&\le &   \sum_i \|c_i \om_i\|  \nonumber\cr
&\le & \|v\| + \eps  \nonumber\ea
Since $\epsilon$ is arbitrary we have $\|Tv\| \le  \|v\|$.
Thus $T$ extends uniquely to a map from $\Tubes$ to $\H$.
\qed

It follows from
Lemmas \ref{lem3'} and \ref{lem4'} that there is a unique homomorphism
from the tube algebra to the holonomy algebra
extending the map $T \maps \Tubes \to \H_c$.
We write this homomorphism as
\[    T \maps S_c\Tubes \to \H_c.\]
This is our final form of the map $T$.
Since $\H_0$ is dense in $\H_c$,
the range of $T$ is dense.  It follows from C*-algebra theory
that $T$ is onto.

Since $S_c\Tubes$ is the completion of $S\Tubes$ in a weaker norm than
$S_b\Tubes$, we have a one-to-one homomorphism with dense range
\[           S_b\Tubes \hookrightarrow S_c\Tubes.\]
We have analogs of Theorems \ref{thm1}
and \ref{thm2} for the tube C*-algebra and holonomy C*-algebra:

\begin{theorem} \label{thm3}\DOT  Let $S_c\Tubes$ be the
tube C*-algebra of $M$, and suppose that $\dim M \ge 3$.
Suppose that $\mu \in (S_c\Tubes)^\ast$ is invariant
under the action of $\Diff_0(M)$.  Then there is a complex-valued
ambient isotopy
invariant $\L_\mu$ of framed oriented links in $M$ given by
\[           \L_\mu(L) = \mu(\Gamma_1 \cdots\Gamma_p) \]
where $\Gamma_1 ,   \cdots , \Gamma_p$ are any choice of normalized
nonintersecting tubes for which $L = L(\Gamma_1, \dots, \Gamma_p)$.
Moreover, the map $\mu \mapsto \L_\mu$ is one-to-one.   \end{theorem}

Proof - This follows from Theorem \ref{thm1} and the fact that
the homomorphism $S_b\Tubes \to S_c\Tubes$ has dense range.  \qed

\begin{theorem} \label{thm4}  \DOT  Let $P$ be a principal $G$-bundle
over a manifold $M$ of dimension $\ge 3$,
 and $\rho$ a finite-dimensional unitary
representation of $G$.  Let $\H_c$ be the holonomy C*-algebra
associated to this data.
The map from $\Diff_0(M)$-invariant elements of
$\H^\ast$ to complex invariants of framed oriented links in $M$
given by
\[           \mu \mapsto \L_{T^\ast \mu}  \]
is one-to-one.  \end{theorem}

Proof - This follows from Theorem \ref{thm2} and the fact that
the homomorphism $\H_b \to \H_c$ has dense range.  \qed

And again, we have open questions:

\begin{question}\DOT \label{q3} What is the range of the map $\mu \mapsto
\L_\mu$ given in Theorem \ref{thm3}?  \end{question}

\begin{question}\DOT \label{q4}
 What is the range of the map $f \mapsto \L_{T^\ast\mu}$ given in
Theorem \ref{thm4}?  \end{question}

We conclude with some remarks on the loop transform.
By the Gelfand-Naimark theorem, the holonomy C*-algebra $\H_c$ is
isomorphic to the algebra of continuous functions on some compact
Hausdorff space $X$. Since any point in $\A/\G$ determines a pure state
on $\H_c$, hence a point of $X$, and since
any element of $\H_c$ that is annihilated by all such pure states must
vanish, we may regard $X$ as a compactification of $\A/\G$, and
write
\[         X = \overline{\A/\G}  .\]
Elements of $\H_c^\ast$ are in one-to-one correspondence with finite
regular Borel measures on $\overline{\A/\G}$.   The map
\[    T^\ast \maps \H_c^\ast \to (S_c\Tubes)^\ast  \]
may be regarded as assigning to each such measure on $\overline{\A/\G}$
its ``loop transform,'' or, since we are working with tubes, its
``tube transform.''  We have shown that this version of the loop
transform is one-to-one.  To further develop the loop representation of
gauge theories it would be useful to have an inverse loop transform.  For
this, one would like an answer to the following:

\begin{question} \DOT \label{q5}
 What is the range of the map $T^\ast \maps \H_c^\ast \to
(S_c\Tubes)^\ast$?  \end{question}

\vfill
\end{document}